\documentclass[twocolumn,aps,prl,showpacs,groupedaddress,amsmath,amssymb]{revtex4}
\usepackage{amsmath}
\usepackage{amssymb}
\usepackage{txfonts}
\usepackage{graphicx}
\begin{document}
\title {Dispersive CQED interactions between matter qubits and bright
squeezed light}
\author{Fang-Yu Hong}
\author{  Shi-Jie Xiong }

\affiliation{National Laboratory of Solid State Microstructures and
Department of Physics, Nanjing University, Nanjing 210093, China}
\date{\today}
\begin{abstract}
Dispersive interactions of matter qubits with bright squeezed light
in a high-Q cavity is studied. Numerical simulation shows that
higher fidelity of operations to obtain a certain phase shift of the
pulse through the dispersive light-matter interaction may be reached
using bright squeezed light than that using bright coherent light.

\end{abstract}

\pacs{03.67.HK,03.67.Mn,42.50.Pq}

\maketitle

The dispersive interaction of an intensive optical pulse with a
single atom in a high-Q cavity has been explored by many experiments
in cavity quantum electrodynamics (CQED) \cite{jrmb}. Such
interactions are essential for non-destructive measurement of atoms
\cite{phbk,rlts,jhjc}, quantum optic computers \cite{tskn}, and
quantum communication \cite{loo}. Ladd {\it et al} \cite{lad} have
studied the interaction between  an intense, off-resonant coherent
optical pulse and a single atom in a high-Q cavity. In this paper,
the cavity-based dispersive interaction of  bright squeezed light
with a three-level atom is been discussed. Numerical simulation
shows that to achieve a certain detectable phase shift of the bright
pulse, higher fidelity of operation may be obtained using squeezed
pulses than that using coherent pulses.

 The basic matter qubit in a cavity is formed by the two
lower states of a three-state $\Lambda$-system, as shown in Fig.1.
The two metastable qubit states are denoted by $|0\rangle$ and
$|1\rangle$. Coherence transitions (rotations) between these two
states are presumed to be possible through methods, such as
stimulated adiabatic Raman transitions \cite{che} or spin-resonance
techniques \cite{feh}. In this article, we focus on optical
transitions between $|1\rangle$ and an excited state $|e\rangle$. We
assume that our light is completely ineffective at inducing
transitions between $|0\rangle$ and $|e\rangle$ either because of
the too far off-resonance of $|e\rangle$, a prohibitive selection
rule, or some combination of the two. One example of such a system
may be found in a semiconductor donor-bound impurity, where the
qubit states are provided by electron Zeeman sublevels and
$|e\rangle$ is provided by the lowest bound-exciton state. In this
paper, the matter qubit is always referred to as an atom although it
may be a semiconductor impurity or quantum dot comprised of many
atoms. Particularly, we assume the state of the qubit is  in the
state $(|0\rangle+|1\rangle)/\sqrt{2}$. The probe pulse is
sufficiently detuned from the transition between $|1\rangle$ and the
exited state to guaranty a strictly weak dispersive light-matter
interaction.

 When large number photons are introduced into a
cavity, a numerical approach is required. For very large photon
numbers, a full-quantum analysis may be computationally intensive;
an appropriate approximation is the semi-classical optical Bloch
equation approach. We presume that $\omega_p=\omega_0$, that is, the
center frequency $\omega_p$ of the pulse is on-resonance with the
cavity (and both are offset from the atomic transition by $\omega_0$
). To keep track of the atomic dynamics, we define several 'partial'
characteristic functions,
\begin{equation}\label{a0}
\chi^{jk}(\eta,t)=\text{Tr}\langle j|exp[\eta
a^\dagger_{in}-\eta^\ast a_{in}]\tilde{\rho}(t)|k\rangle,
\end{equation}
where $\tilde{\rho}(t)$ is the density operator of the light-matter
system, states $|j\rangle$ and $|k\rangle$ are atomic states and the
trace is over the optical field. Assuming a rotating reference frame
rotating at the center frequency of the optical pulse, for a
narrow-band pulse, the master equations in a fully quantum setting
in which any quantum state of light is allowed have the form
\cite{lad}:

\begin{widetext}
\begin{subequations}\label{a}
\begin{eqnarray}
\dot{\chi}^{ee}(\eta,t)&=&ig\left[S(t)\left(\frac{\eta}{2}+\frac{\partial}{\partial\eta^\ast}\right)\,\chi^{1e}(\eta,t)+
S^\ast(t)\left(\frac{\eta^\ast}{2}+\frac{\partial}{\partial\eta}\right)\,\chi^{e1}(\eta,t)\right]-2\Gamma\,\chi^{ee}(\eta,t),
\label{a1}
\\
\dot{\chi}^{11}(\eta,t)&=&ig\left[S(t)\left(\frac{\eta}{2}-\frac{\partial}{\partial\eta^\ast}\right)\,\chi^{1e}(\eta,t)+
S^\ast(t)\left(\frac{\eta^\ast}{2}-\frac{\partial}{\partial\eta}\right)\,\chi^{e1}(\eta,t)\right]+2\Gamma\,\chi^{ee}(\eta,t),
\label{a2}
\\
\dot{\chi}^{e1}(\eta,t)&=&igS(t)\left[\left(\frac{\eta}{2}-\frac{\partial}{\partial\eta^\ast}\right)\,\chi^{ee}(\eta,t)+
\left(\frac{\eta^\ast}{2}+\frac{\partial}{\partial\eta}\right)\,\chi^{11}(\eta,t)\right]+
(i\Omega-\Gamma)\,\chi^{ee}(\eta,t).
\label{a3}\\
\dot{\chi}^{e0}(\eta,t)&=&igS(t)\left(\frac{\eta}{2}+\frac{\partial}{\partial\eta^\ast}\right)\,\chi^{10}(\eta,t)+
(i\Omega-i\Delta-\Gamma)\,\chi^{e0}(\eta,t),\label{a4}\\
\dot{\chi}^{10}(\eta,t)&=&igS(t)^\ast\left(\frac{\eta^\ast}{2}-\frac{\partial}{\partial\eta}\right)\,\chi^{e0}(\eta,t)
-i\Delta\,\chi^{10}(\eta,t)\label{a5},\\
\dot{\chi}^{00}(\eta,t)&=&0\label{a5a},
\end{eqnarray}
\end{subequations}
\end{widetext}
where $g$ is the atom-cavity coupling factor, $S(t)$ is related to a
cavity-waveguide coupling factor $\kappa$, cavity decay parameter
$\gamma$ which imply that any optical power in the cavity leaks out
of the cavity as $e^{-\gamma t}$, and the input pulse shape
$S_{in}(t)$ coupling into the cavity as follows\cite{lad},
\begin{equation}\label{eqs}
S(t)=2\sqrt{\kappa}\frac{S_{in}(t)}{\gamma}
\end{equation}
 In Eq.\eqref{a}, $2\Gamma$ is the total decay rate of the atom in the
cavity, including the influence of the Purcell effect:
\begin{equation}
2\Gamma=\frac{1+P(\omega_p)}{\tau_r}+\frac{1}{\tau_{nr}},
\end{equation}
where $\tau_r$ and $\tau_{nr}$ describes spontaneous emission and
non-radiative decay, respectively, and $P(\omega)$ is the Purcell
factor
\begin{equation}
P(\omega)=\frac{\tau_r\gamma g^2}{\omega^2+\gamma^2/4}.
\end{equation}
In Eq.\eqref{a}, $\Omega$ is the atomic detuning from the cavity,
including the ac-Stark shift,
 \begin{equation}
 \Omega=\omega_p\left[1+\frac{P(\omega_p)}{\gamma \tau_r}\right].
 \end{equation}

\begin{figure}
\includegraphics[scale=0.24]{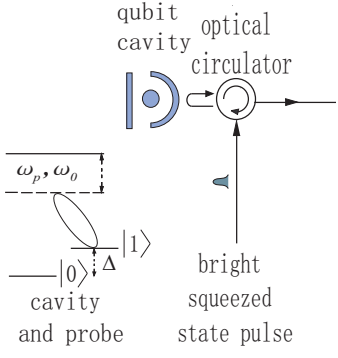}
\caption{\label{fig:1} Schematic of the dispersive interaction
between a three-level atom and a bright squeezed pulses in a high-Q
cavity.}
\end{figure}

Numerical solutions of this system of equations at large number
photons are computationally intensive, so we use the semiclassical
approximation. The assumptions underpinning the semiclassical
approximation are that the quantum state of the pulse during the
light-matter interaction is always a squeezed state, and that it
always remains unentangled from state $|e\rangle$. (Similar
assumptions were used in \cite{lad}.) Then the density operator has
the form
\begin{eqnarray}\label{a6}
\tilde\rho(t)&=&|\,\tilde\beta(t)\rangle_{gg}\langle\,\tilde\beta(t)|\otimes
\{\,\rho^{ee}(t)\sigma^+\sigma^-+[\,\rho^{11}(0)-\rho^{ee}(t)]\sigma^-\sigma^+\notag\\
&+&\rho^{e1}(t)\sigma^++\rho^{1e}(t)\sigma^-\}
+|\,\tilde\beta(t)\rangle_{gg}\langle\,\beta|\otimes\{\,\rho^{e0}(t)|e\rangle\langle
0|\notag\\
&+&\rho^{10}(t)|1\rangle\langle0|\}
+|\,\beta\rangle_{gg}\langle\,\tilde\beta(t)|\otimes\{\,\rho^{0e}(t)|0\rangle\langle
e|+\rho^{01}(t)|0\rangle\langle1|\}\notag\\
&+&|\,\beta\rangle_{gg}\langle\,\beta|\otimes\rho^{00}|0\rangle\langle0|,
\end{eqnarray}
where $\sigma^+=|e\rangle\langle1|$, $\sigma^-=|1\rangle\langle e|$,
and $|\beta\rangle_g$ is a two-photon coherence state defined as
\begin{equation}
|\beta\rangle_g=\Xi(\varepsilon)D(\beta)|0\rangle,
\end{equation}
where
\begin{equation}
D(\beta)=\exp(\beta a^\dagger-\beta^* a),
\end{equation}
and
\begin{equation}
\Xi(\varepsilon)=\exp(\frac{1}{2}\varepsilon^*a^2-\frac{1}{2}\varepsilon
a^{\dagger2})
\end{equation}
are the displacement operator and the unitary squeezed operator
respectively, $a^\dagger$ is the creation operator of photons and
$\varepsilon=re^{2i\phi}$, $r$ is the squeeze factor. Squeezed
states $|\alpha,\varepsilon\rangle$ defined by
\begin{equation}
|\alpha,\varepsilon\rangle=D(\alpha)\Xi(\varepsilon)|0\rangle,
\end{equation}
are equivalent to two-photon coherence states $|\beta\rangle_g$
\cite{wal}:
\begin{equation}
|\alpha,\varepsilon\rangle=|\beta\rangle_g,
\end{equation}
where
\begin{equation}\label{eq2}
\beta=\mu\alpha+\nu\alpha^*,
\end{equation}
with $\mu=\cosh r$ and $\nu=e^{2i\phi}\sinh r$.

Substituting this density operator  into equations
(\ref{a0}-\ref{a5a}), focusing on $\eta=0$, using the following
formulas \cite{wal}
\begin{equation}
\Xi^\dagger(\varepsilon)\, a\,\Xi(\varepsilon)=\mu a-\nu^\ast
a^\dagger ,
\end{equation}

\begin{equation}
\Xi^\dagger(\varepsilon)\,a^\dagger\, \Xi(\varepsilon)=\mu
a^\dagger-\nu a ,
\end{equation}
we arrive at the optical Bloch equations:
\begin{subequations}\label{b}
\begin{eqnarray}
\dot\rho^{ee}&=&ig[S^\ast(t)\tilde\delta(t)^\ast\rho^{e1}(t)-
S(t)\tilde\delta(t)\rho^{1e}(t)]-2\Gamma\rho^{ee}, \label{b1}\\
\dot\rho^{e1}&=&igS(t)\tilde\delta(t)[2\rho^{ee}(t)-
\rho^{11}(0)]+(i\Omega-\Gamma)\rho^{e1},\label{b2}\\
\dot\delta&=&-igS^\ast(t)\rho^{e1}(t)\frac{1+\mu^2+|\nu|^2}{2(\rho^{11}(0)-\rho^{ee}(t))},\label{b3}\\
\dot\rho^{e0}&=&-igS(t)\tilde\delta(t)\rho^{10}(t)\notag
\\&-&[i\Delta-i\Omega+\Gamma+c(t)]\rho^{e0}(t),\label{b4}
\\
\dot\rho^{10}&=&-igS^\ast(t)\delta^\ast\rho^{e0}(t)-
[i\Delta+c(t)]\rho^{10}(t),\label{b5}\\
\dot\rho^{00}&=&0,\label{b6}
\end{eqnarray}
\end{subequations}
where
\begin{equation}
c(t)=\frac{\partial}{\partial
t}\ln\langle\beta|\tilde\beta(t)\rangle=-\frac{1}{2}\frac{\partial}{\partial
t}|\tilde\beta(t)|^2+\beta^\ast\dot{\tilde\beta}(t),
\end{equation}
with
\begin{equation}\label{a7}
    \tilde{\beta}(t)=\mu\tilde{\alpha}(t)+\nu
    \tilde{\alpha}(t)^\ast,
\end{equation}
 $\Delta$ is the energy separation of states $|0\rangle$ and
$|1\rangle$, and
\begin{equation}
\tilde\delta(t)=\mu\tilde\beta(t)-\nu\tilde\beta^\ast(t)=\tilde{\alpha}(t).
\end{equation}

If $\alpha$ experiences  a phase shift of $\theta$, i.e.,
$a(t)=ae^{i\theta}$, according to the following formula \cite{mors}:
\begin{eqnarray}
\langle \Delta N(t)\rangle^2&=&\langle \Delta
N(0)\rangle^2=2\sinh^2r\cosh^2r\notag\\
&+&|\alpha|^2\left[e^{-2r}\cos^2(\theta-\frac{\phi}{2})+e^{2r}\sin^2(\theta-\frac{\phi}{2})\right],
\end{eqnarray}
where $N=a^\dagger a$ and $\langle\,\rangle$ denotes the expectation
value for squeezed state $|\alpha,\varepsilon\rangle$, $\varepsilon$
should have a corresponding phase shift of $2\theta$.
 For simplicity,  hereafter, we
assume $\alpha$ and $\varepsilon$ are real.

\begin{figure}
\includegraphics[scale=0.4]{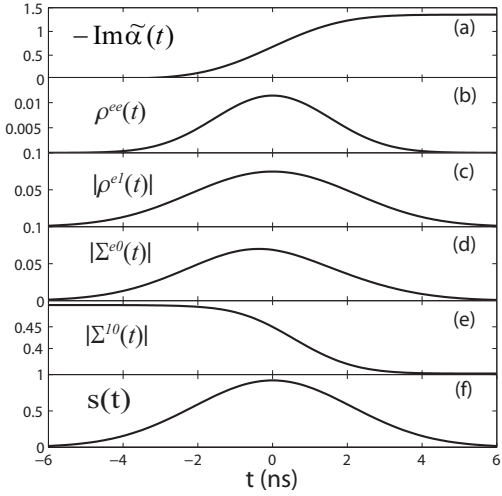}
\caption{\label{fig:2} Numerical simulations of the weak dispersive
interaction between squeezed lights and a matter qubit in a cavity.
The parameters are $\alpha=100$, $r=1$, $g/2\pi=0.17$ GHz,
$\Gamma/2\pi=1$ MHz, $\kappa=\gamma=0.2\times 2\pi$ GHz,
$\sigma=3$ns, and $\Omega/2\pi=100$ GHz.}
\end{figure}
\begin{figure}

\includegraphics[scale=0.4]{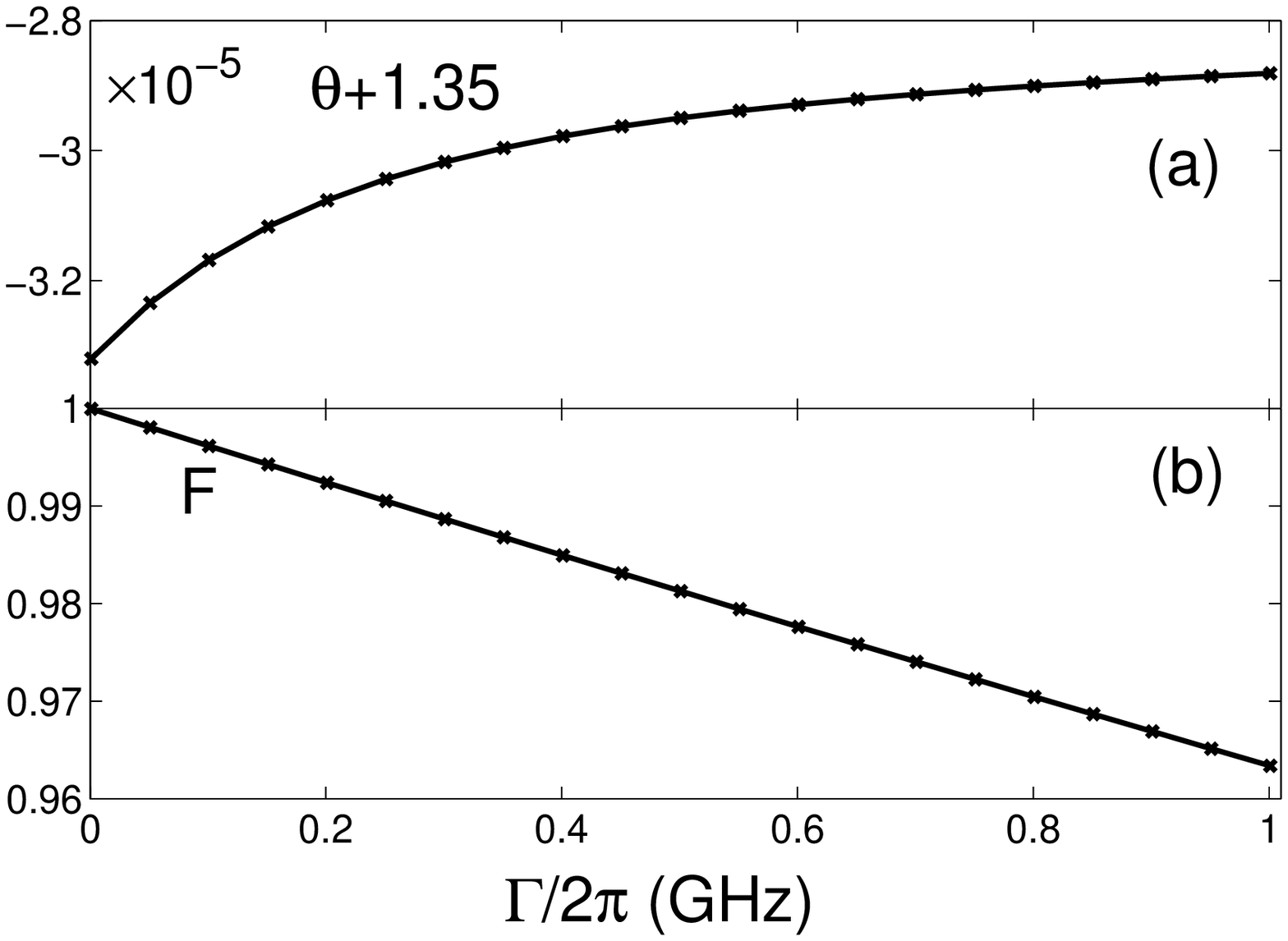}
\caption{\label{fig:3}The phase shift $\theta$ of
$\tilde{\alpha}(t)$ and fidelity $F$ of the matter qubit after
dispersive light-matter interaction versus $\Gamma$.The parameters
are $\alpha=100$, $r=1$, $g/2\pi=0.17$ GHz, $\kappa=\gamma=0.2\times
2\pi$ GHz, $\sigma=3$ns, and $\Omega/2\pi=100$ GHz.}
\end{figure}

\begin{figure}
\includegraphics[scale=0.45]{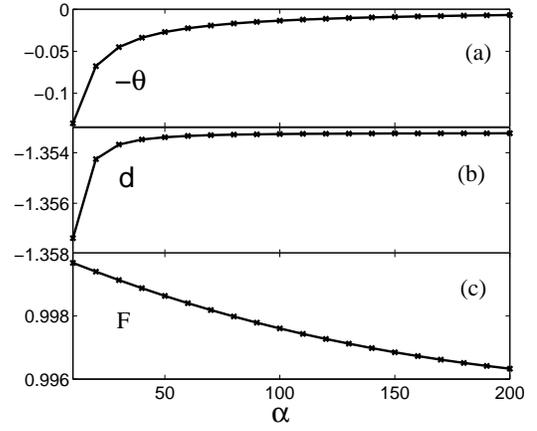}
\caption{\label{fig:4} The phase shift $\theta$, the
distinguishability $d$, and the  fidelity $F$ of the matter qubit
after dispersive light-matter interaction versus $\alpha$. The
parameters are $r=1$, $g/2\pi=0.17$ GHz, $\kappa=\gamma=0.2\times
2\pi$ GHz, $\sigma=3$ ns, and $\Omega/2\pi=100$ GHz.
$\Gamma/2\pi=10$ MHz.}
\end{figure}

\begin{figure}
\includegraphics[scale=0.5]{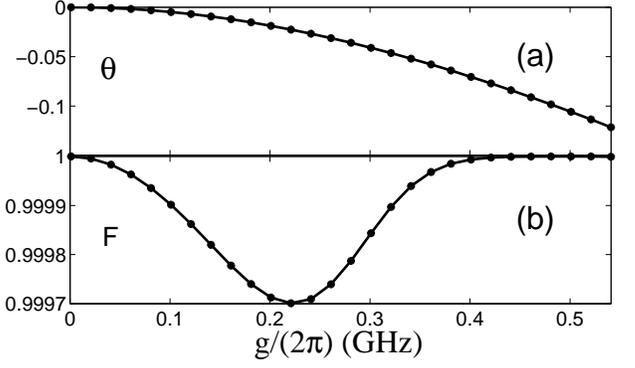}
\caption{\label{fig:5} The phase shift $\theta$ of
$\tilde{\alpha}(t)$ and the fidelity $F$ of the matter qubit after
dispersive light-matter interaction versus coupling $g$.The
parameters are $\alpha=100$, $r=1$, $\Gamma/2\pi=1$MHz,
$\kappa=\gamma=0.2\times 2\pi$ GHz, $\sigma=3$ ns, and
$\Omega/2\pi=100$ GHz. }
\end{figure}

\begin{figure}
\includegraphics[scale=0.4]{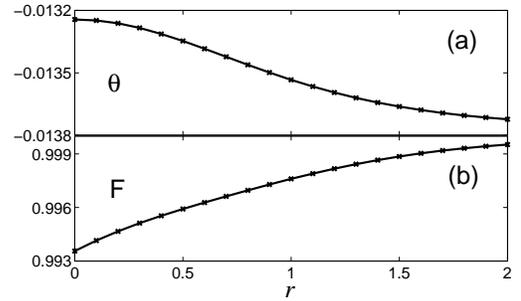}
\caption{\label{fig:6} The phase shift $\theta$ of
$\tilde{\alpha}(t)$ and the fidelity $F$ of the matter qubit after
dispersive light-matter interaction versus squeeze factor $r$.The
parameters are $\alpha=100$,
$g/(2\pi)=(\cosh^2(1)+\sinh^2(1))/(\cosh^2(r)+\sinh^2(r))\times0.17$
GHz, $\Gamma/2\pi=10$MHz, $\kappa=\gamma=0.2\times 2\pi$ GHz,
$\sigma=3$ ns, and $\Omega/2\pi=100$  GHz. }
\end{figure}

\begin{figure}
\includegraphics[scale=0.4]{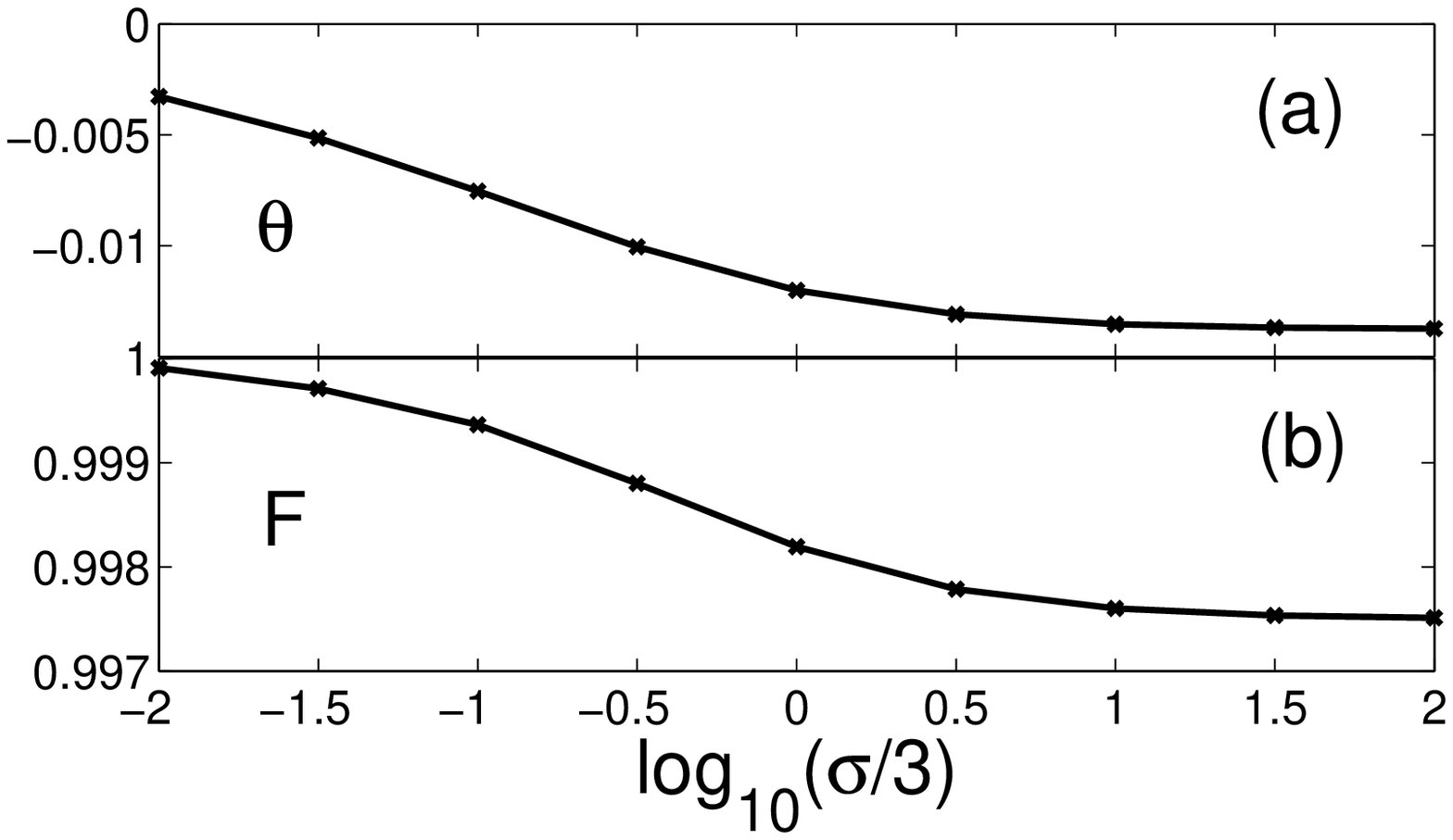}
\caption{\label{fig:7} The phase shift $\theta$ of
$\tilde{\alpha}(t)$ and the fidelity $F$ of the matter qubit after
dispersive light-matter interaction versus $\log_{10}(\sigma/3)$
where $\sigma$ is in units of ns.The parameters are $\alpha=100$,
$r=1$, $g/2\pi=0.17$ GHz, $\Gamma/2\pi=10$ MHz,
$\kappa=\gamma=0.2\times 2\pi$ GHz,  and $\Omega/2\pi=100$ GHz. }
\end{figure}

\begin{figure}
\includegraphics[scale=0.45]{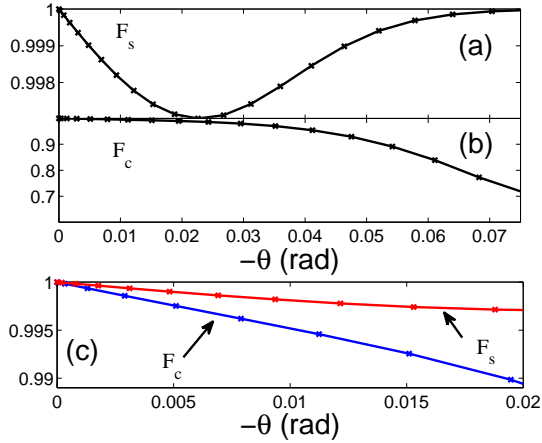}
\caption{\label{fig:8} The fidelity of the matter qubit after
dispersive light-matter interaction $F_s$ using squeezed light and
$F_c$ using coherent light versus phase shift $-\theta$ of
$\tilde{\alpha}(t)$.The parameters are $\alpha=100$,
$\Gamma/2\pi=10$MHz, $\kappa=\gamma=0.2\times 2\pi$ GHz,
$\Omega/2\pi=100$ GHz, $r=1$ (a), and $r=0$ (b).}
\end{figure}

We first discuss the approximate solution of Eq.\eqref{b}, then
numerically solve them. Eq.\eqref{b4},\eqref{b5} are related to the
fidelity of dispersive light-matter interaction, which is degraded
by internal loss. Those two equations show a phase advancement by
$\Delta$, and both a phase and loss from $c(t)$, which corresponds
to the phase advance and dephasing from the light. They can be
simplified by define
\begin{subequations}
\begin{eqnarray}
\Sigma^{e0}(t)=\langle\,\beta\,|\,\tilde\beta(t)\rangle
e^{i\Delta}\rho^{e0}(t)\label{c1},\\
\Sigma^{10}(t)=\langle\,\beta\,|\,\tilde\beta(t)\rangle
e^{i\Delta}\rho^{10}(t)\label{c2},
\end{eqnarray}
\end{subequations}
which obey
\begin{subequations}
\begin{eqnarray}
\dot\Sigma^{e0}(t)&=&-ig\tilde\delta(t)S(t)\Sigma^{10}(t)+(i\Omega-\Gamma)\Sigma^{e0}(t),\label{c3}\\
\dot\Sigma^{10}(t)&=&-ig\delta^\ast
S^\ast(t)\Sigma^{e0}(t)\label{c4}.
\end{eqnarray}
\end{subequations}

 From Eq.\eqref{b1},\eqref{b3}, we obtain
\begin{equation}
\begin{split}
|\tilde\alpha(t)|^2&=|\alpha|^2-\int_0^{\,t}\dot\rho^{ee}(t')
\frac{1+\mu^2+|\nu|^2}{2\rho^{11}(0)-2\rho^{ee}(t)}dt'\\
&-2\Gamma\int_0^{\,t}\rho^{ee}(t')\frac{1+\mu^2+|\nu|^2}{2\rho^{11}(0)-2\rho^{ee}(t)}dt'
\end{split}
\end{equation}
 Since $\rho^{ee}(t)\ll\rho^{11}(0)$, which we will see in the
 following numerical simulation, and $\rho_a^{ee}(t)\longrightarrow0$
 as $t\longrightarrow\infty$, we arrive at
 \begin{equation}
|\tilde\alpha(\infty)|^2=|\alpha|^2-2\Gamma\int_0^\infty\rho^{ee}(t')\frac{1+\mu^2+
|\nu|^2}{2\rho^{11}(0)}dt'.
 \end{equation}

All optical losses arise finally from atomic decay. Other optical
losses from the cavity independent from atomic decay are already
incorporated into the definition of $S(t)$. We may approximately
solve Eq.\eqref{b1}-\eqref{b3} for the phase shift and optical loss
in the limit of a narrow-band, far detuned pulse. This approximation
may be obtained by assuming $gS(t)\tilde\delta(t)$ is constant in
time, with value $g\bar{S}(t)\alpha$, and solving them for
$\rho^{ee}$ and $\rho^{e1}$. (A similar approach was adopted by
\cite{lad} and \cite{hor}.) Using Laplace transforms, only the
zero-valued pole of $\rho^{ee}$ and
 $\rho^{e1}$ being important, we may arrive at the approximation
 solutions
\begin{equation}
\rho^{ee}\rightarrow
\frac{\rho^{11}(0)g^2|S(t)\alpha|^2}{\Gamma^2+\Omega^2+2g^2|\bar{S}\alpha|^2},
\end{equation}
\begin{equation}
\rho^{e1}\rightarrow
\frac{\rho^{11}(0)gS(t)\alpha}{\Gamma^2+\Omega^2+2g^2|\bar{S}\alpha|^2}(\Omega-i\Gamma).
\end{equation}
Presuming this solution for $\rho^{e1}$ is maintained as $S(t)$
varies in time, integrating Eq.\eqref{b3}, we find
\begin{equation}\label{eq20}
\tilde\alpha(t)=\alpha\left[1-ig^2\int_0^tdt'|S(t')|^2\frac{(\Omega-i\Gamma)(1+\mu^2+\nu^2)}
{2(\Gamma^2+\Omega^2+2g^2|\bar{S}\alpha|^2)} \right]
\end{equation}
In a same way, we can find approximation solution of
Eq.\eqref{c3},\eqref{c4},
\begin{eqnarray}
\Sigma^{e0}&\rightarrow&\Sigma^{e0}(0)\frac{ig\tilde\alpha(t)S(t)}{2(i\Omega-\Gamma)},\label{eq21}\\
\Sigma^{10}&\rightarrow&\Sigma^{10}(0)\exp\left(-g^2\int_0^\infty
\frac{\alpha\ast\tilde\alpha(t)|S(t')|^2}{2(\Gamma-i\Omega)}\right)dt'\label{eq22}.
\end{eqnarray}
From Eq.\eqref{eq20},\eqref{eq22}, we may find that to achieve a
certain phase shift of $\alpha$, the larger the squeeze factor $r$,
the smaller the magnitude of $g$, thus the higher the fidelity of
the matter qubit after dispersive light-matter interactions. The
total magnitude of the damping to the desired coherence is
\begin{equation}
\left|\rho^{10}(t)\right|=
e^{|\beta-\tilde\beta(t)|^2/2}\left|\Sigma^{10}(t)\right|
\end{equation}
For the calculations presented here, we assume this interaction is
used for entanglement distribution, in which case $\rho^{10}=1/2$.
Then the final fidelity of our qubit may be written
\begin{equation}
F=\frac{1}{2}(1+2|\rho^{10}(t)|).
\end{equation}

Now we discuss the numerical solution of equations
\eqref{b1}-\eqref{b3},\eqref{c3},\eqref{c4}. All the following
simulations assume that $S_{in}$ takes a Gaussian shape,
$S_{in}=\sqrt{\frac{\sqrt 2}{\sqrt \pi
\sigma_p}}\exp(-\frac{t^2}{\sigma_p^2})$, and $\kappa=\gamma$, thus,
from Eq.\eqref{eqs}, we have
\begin{equation}
S(t)=2\sqrt{\frac{\sqrt 2}{\gamma\sqrt {\pi}
\sigma}}\exp(-\frac{t^2}{\sigma^2}).
\end{equation}
We also presume $ \varepsilon=re^{i\pi}$ and  the initial state of
the matter qubit is $(|0\rangle+|1\rangle)/2$. The parameters  are
assumed to be $\Omega/2\pi=$ 100GHz, $\kappa/2\pi=\gamma/2\pi=0.2$
GHz, $g/2\pi=0.17$ GHz, which are typical for $^{31}P$ \cite{lad} ,
and $\alpha=10$, $r=0$ for coherent state, $|\alpha\rangle$,
  we have (1)
$F_r=0.99999724$,  $\theta=-5.77896$,
$\tilde{\alpha}(\infty)-|\alpha|=-5.8\times10^{-7}$ for
$\Gamma/2\pi=1$ MHz, and (2)$F_i=1.00000014$, $\theta=-5.77896$,
$\tilde{\alpha}(\infty)-|\alpha|=2.7\times10^{-13}$ for
$\Gamma/2\pi=0$. $F>1$ shows that the  equations (\ref{c3},
\ref{c4}) does not accurately describe the fidelity of the
dispersive interaction. This problem may be solved by including
higher order terms in the equations. Here we overcome this problem
in a simple way: because under the ideal situation where
$\Gamma/2\pi=0$ and $\kappa=\gamma$, there will be no decoherence
arising from the dispersive interaction and the fidelity $F_i$
should be unity, we may use $F_i$ obtained from $\Gamma=0$ with
other parameters being the same as those used to obtain the fidelity
$F_r$ for $\Gamma\neq0$ as the fidelity reference, i.e. we hereafter
calculate fidelity using the formula $F=F_r/F_i$.  In this way, for
the aforesaid example, we have fidelity $F=0.999997$.

 The numerical simulations are
shown in Fig. \ref{fig:2}. From the simulation, we find that
$|\alpha-|\tilde\alpha(\infty)||/\alpha=1.3\times10^{-7}$, which
show that the change in the magnitude of $\alpha$ is absolutely
negligible, and so does the absorption of photons in the
interaction. The fidelity of the matter qubit of this operation is
$\zeta=0.9998$, with $\rho_a^{10}=0.4997\equiv0.5 \zeta$ with
$\zeta=0.9994$. The phase shift of $\tilde{\alpha}(t)$ is
$\theta=-0.01353$. Figure \ref{fig:3} shows that the magnitude of
the phase shift $-\theta$ decrease very slight, so does the fidelity
$F$, as the atomic decay $\Gamma$ increase in three orders from
$10^{-3}$ to 1. Figure \ref{fig:4} shows the dependence of the phase
shift $\theta$, the distinguishability
$d\equiv\tilde{\alpha}(\infty)\sin\theta$, and the fidelity $F$  on
the magnitude of $\alpha$. With the increasing of the magnitude of
coupling factor $g$, the fidelity $F$ decreases at first, then
increases again (see Fig. \ref{fig:5}), which shows that the
decoherence factor $\Gamma$ plays less role when $g$ become larger
than about $0.22\times2\pi$ GHz.

If we increase the squeeze factor $r$ while keeping
$g(\cosh^2r+\sinh^2r)$ constant, the fidelity $F$ and the magnitude
of phase shift $\theta$ both increase (see Fig. \ref{fig:6}). The
phase $\theta$ and the fidelity $F$ are dependent on the length of
the pulse $\sigma$ (see Fig. \ref{fig:7}). The results of Fig.
\ref{fig:8} tell us that we can obtain higher fidelity of the
dispersive interaction  using squeezed pulses to get a certain phase
shift of $\tilde{\alpha}(t)$ than that using coherent ones. Those
characters show that this scheme may have good adaptability to wide
range different systems.

In conclusion, this paper has discussed  the dispersive interaction
of bright squeezed light with an three-level atom in a high-Q
cavity. Numerical simulation shows that (1) the lower decoherence of
the atom arising from the interaction with the light will available,
the larger the squeeze factor of the squeezed pulse is, (2) compared
with that using bright coherent light, higher fidelity of the atom
qubit can be realized using bright squeezed light.

 {\it Acknowledgments} This work was supported by National Foundation of
Natural Science in China Grant Nos. 60676056 and 10474033, and by
the China State Key Projects of Basic Research (2006CB0L1000 and
2005CB623605).


\begin{references}
\bibitem{jrmb} J.M. Raimond, M. Brune, and S. Haroche, Rev. Mod. Phys. {\bf73}, 565 (2001).
\bibitem{phbk} P. Horak, B.G. Klappauf, A. Haase, R. Folman, J. Schmiedmayer, P. Domokos, and E.A. Hinds, Phys. Rev. A {\bf67}, 43806 (2003).
\bibitem{rlts} R. Long, T. Steinmets, P. Hommelhoff, W. H\"{a}nsel, T.W. H\"{a}nsch, and J. Reichel, Phil. Trans. R. Soc. Lond. A {\bf361}, 1375 (2003).
\bibitem{jhjc} J.J. Hope and J.D. Close, Phys. Rev. Lett. {\bf
93},180402 (2004).
\bibitem{tskn} T.P. Spiller, K. Nemoto, S.L. Braunstein, W.J. Munro, P. van Loock, and G.J. Milburn, New J.Phys. {\bf 8},30 (2006).
\bibitem{loo}P.van Loock, T.D.Ladd, K.Sanaka, F.Yamaguchi, K.Nemoto, W.J.Munro,
and Y.Yamamoto, Phys. Rev. Lett. {\bf 96},240501 (2006).
\bibitem{lad}T.D.Ladd, P.van Loock, K.Nemoto, W.J.Munro, and Y.Yamamoto, New J.Phys. {\bf 8},184 (2006).
\bibitem{che}P.Chen, C.Piermarocchi, L.J.Sham, D.Gammon, and D.G.Steel,  Phys. Rev. B {\bf 96}, 010502 (2006).
\bibitem{feh}G.Feher, Phys. Rev. {\bf 114}, 1219 (1959).
\bibitem{wal}D.F.Wall, and G.J.Milburn 1995 Quantum Optics (Berlin: Springer-Verlag).
\bibitem{hor}P.Horak, B.G.Klappauf, A.Haase, R.Folman, J.Schmiedmayer, P.Domokos, and E.A.Hinds,
 Phys. Rev. A {\bf 67}, 43806 (2003).
\bibitem{mors} M. Orszag 2000 Quantum Optics (Berlin: Springer-Verlag).

\end{references}
\end{document}